\documentclass[useAMS,usenatbib]{mn2e}
\usepackage{epsfig}
\usepackage{rotating}
\begin{document}

\title[{\sl {\bf XMM-Newton}} observations of warm absorbers in PG quasars]
    {{\sl XMM-Newton} observations of warm absorbers in PG quasars}
\author[C.E.~Ashton et al.]
    {C.E.~Ashton$^1$\thanks{email: cea@mssl.ucl.ac.uk}, M.J.~Page$^1$, A.J.~Blustin$^1$, E.M.~Puchnarewicz$^1$, 
\newauthor G.~Branduardi-Raymont$^1$, K.O.~Mason$^1$, F.A.~C\'{o}rdova$^2$, W.C.~Priedhorsky$^3$    \\
$^1$Mullard Space Science Laboratory, University College London, Holmbury St. Mary, Dorking, Surrey RH5 6NT\\
$^2$Department of Physics, University of California, Riverside, CA, 92521, USA \\
$^3$Los Alamos National Laboratory, Los Alamos, NM 87545, USA \\}
\date{Accepted ??. Received ??}
\pubyear{2004}
\maketitle

\begin{abstract}

We present {\sl XMM-Newton EPIC} observations of warm absorbers in the quasars PG~1114+445 and PG~1309+355, both of which exhibit evidence for absorption by warm material in the line-of-sight. We find the absorption in PG~1114+445 to be in two phases, a `hot' phase with a log ionisation parameter $\xi$ of 2.57, and a `cooler' phase with log $\xi$ of 0.83; an unresolved transition array (UTA) of M-shell iron is observed in the cooler phase. The absorption in this quasar is similar to that observed in the Seyfert 1 NGC 3783. The absorption in PG~1309+355 consists of a single phase, with log $\xi$ of 1.87. The absorbing gas lies at distances of 10$^{18}$ - 10$^{22}$ cm from the continuum radiation sources in these AGN, which suggests that it could originate in a wind emanating from a molecular torus. We derive distances assuming that these X-ray warm absorbers have the same velocity as the UV warm absorbers observed in these quasars. The distances to the warm absorbers from the central continuum source scale approximately with the square root of the AGN ionising luminosity, a result consistent with the warm absorber originating as a torus wind. The kinetic luminosities of these outflowing absorbers represent insignificant fractions ($< 10^{-3}$) of the energy budgets of the AGN.

\end{abstract}


\section{Introduction}

Warm absorbers are clouds of ionised gas within AGN. They cause absorption at soft X-ray wavelengths and were first identified by Halpern (1984). Photoelectric and line absorption, in gas that lies along our line-of-sight to the nucleus, is an important source of information on how the central engines of quasars affect their immediate environments. {\sl Chandra LETGS} and {\sl XMM-Newton RGS} data have shown that warm absorbers give rise to a series of discrete narrow absorption lines in soft X-rays (e.g. NGC 5548, Kaastra et al 2000; IRAS 13349+2438, Sako et al 2001) and are usually flowing outward from the AGN at a few hundred km s$^{-1}$. The absorption lines provide valuable constraints on basic properties such as outflow velocities, ionisation parameters and column densities.
 
At soft X-ray energies, inner-shell absorption by M-shell iron ions in the form of unresolved transition arrays (UTAs) are apparent in some Seyfert 1 spectra (e.g. Sako et al 2001). UTAs are the main spectral sign of a `cold' phase of the warm absorber, and their position in a spectrum can be used to determine the ionisation state of the gas (Behar et al 2001). 

The origins of warm absorbers are not firmly established.  Explanations include accretion disc winds (Elvis 2000), or a wind resulting from evaporation of the inner torus (Krolik \& Kriss, 2001). In the case of NGC 4151, Crenshaw et al (2000) suggest an association with radiation driven clouds that form the NLR. Absorption by both a torus wind and a disc wind are invoked as possible explanations by Kriss et al (2003) for NGC 7469. 

Of a large sample of Seyfert 1s observed with {\sl ASCA}, $\sim$ 50\% have absorption features in their soft X-ray spectra due to ionised gas (Reynolds 1997). Intrinsic absorption could be universal if the absorbing gas fills only part of the sky as seen from the central continuum source. However, significant X-ray warm absorption appears to be rare in quasars (Laor et al 1997; hereafter L97).

\begin{table*}
\caption{Galactic column densities and observation details.}
\begin{tabular}{ccccccccc}
\hline
Quasar & Date Observed & Redshift & Galactic N$_{\mbox{\sc h}}$ & Effective Exposure Time & Count Rate \\
&&&($10^{20}$cm$^{-2}$)&(ks)&(Counts s$^{-1}$) \\
\hline
PG~1114+445 & 15.05.2002 & 0.144 & 1.83 & PN:37, MOS:42 & PN:0.67, MOS:0.20  \\
PG~1309+355 & 10.06.2002 & 0.184 & 1.03 & PN:23, MOS:29 & PN:0.50, MOS:0.13  \\
\hline
\end{tabular}
\label{quasars}
\end{table*}

In our study of UV-excess selected quasars (Brocksopp et al, in preparation) from the Palomar-Green sample (Schmidt \& Green 1983), two, PG~1114+445 and PG~1309+355, show evidence of such absorption features. In this paper, we present a detailed analysis of these features.

The presence of a warm absorber in PG~1114+445 has been noted before (George et al 1997, L97). A joint analysis of {\sl ROSAT} and {\sl ASCA} data by George et al (1997) found that PG~1114+445 can be modelled well by a power law absorbed by photoionised material, with a column density of 2$\times10^{22}$cm$^{-2}$. In UV spectra, Mathur, Wilkes \& Elvis (1998; hereafter M98) found an absorber which is outflowing with a line-of-sight velocity of $\sim$ 530 km s$^{-1}$. M98 proposed that the X-ray and UV absorption originate in the same material, because it is rare to see either X-ray or UV absorbers in a radio-quiet quasar.

With high signal-to-noise data from the {\sl XMM-Newton EPIC} cameras, much higher quality spectra are obtained than was possible with previous missions. The absorption features in PG~1309+355 were not detected in X-ray observations prior to {\sl XMM-Newton}. 

Recently Porquet et al (2004; hereafter PO4) have modelled {\sl EPIC} data of PG~1114+445 and PG~1309+355 with a simple combination of absorption edges; we compare our and their results in Section.~\ref{Discussion}.

The {\sl EPIC} data were taken in large window mode, and using the thin filter. They were processed using the Science Analysis System (SAS) Version 5.4.1; the {\sl pn} spectra were constructed using single and double events, corresponding to pattern values of 0-4, and the {\sl MOS} spectra were constructed using all valid events (PATTERN = 0-12). The count rates are well below the thresholds (12 counts s$^{-1}$ for the {\sl pn} and 1.8 counts s$^{-1}$ for the {\sl MOS}) where pile-up has to be considered for large window mode. Events next to bad pixels and next to the edges of CCDs were excluded (FLAG = 0 in SAS). The source spectra were constructed by taking counts within a circle radius 25 arcsec around PG~1114+445 and 30 arcsec around PG~1309+355. In each case, background spectra were extracted from nearby source-free regions with three times the radius of those used to extract the source spectra. We excluded periods of high background, which we identified in 5-10 keV lightcurves for the whole field of view outside the source circle. The quasar spectra were grouped with a minimum of twenty counts per channel, to allow $\chi^2$ statistics to be used. Response matrix files (RMF) and auxiliary response files (ARF) for each instrument were generated using the {\sl SAS}. The {\sl EPIC MOS} and {\sl pn} data were coadded using the method of Page, Davis \& Salvi (2003). The data were modelled using {\sl SPEX} 2.00 (Kaastra et al 2002). In this paper we adopt the values H$_{0}$ = 70.0 km s$^{-1}$ Mpc$^{-1}$, $\Omega_{m}$ = 0.3 and $\Omega_{\lambda}$ = 0.7.

\section{Results}
\label{Results}

\subsection{Initial fitting}

\begin{figure*}
\epsfig{file=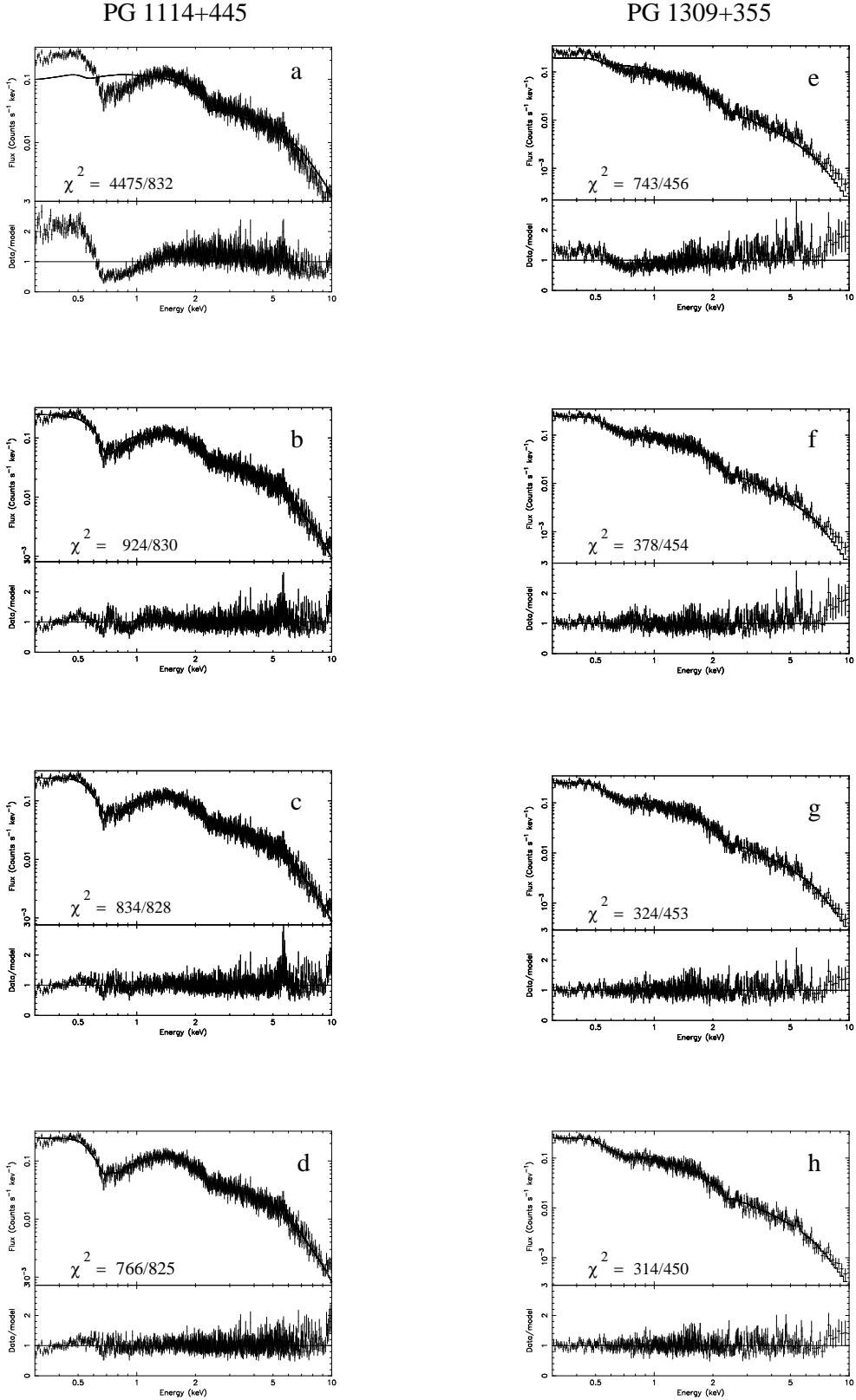,width=13cm,angle=0}
\caption{Spectral fits of combined {\sl MOS and pn} data, plotted in the observed frame on logarithmic scales. The plots below each fit show the data/model ratio. The $\chi^{2}$/d.o.f values are shown for each fit. Left-hand column: PG~1114+445. a) Power law fit, b) fit with single warm absorber, c) fit with two warm absorbers, d), same as c) but with Fe K$\alpha$ line. Right hand column: PG~1309+355. e) Power law fit, f) single warm absorber fit without blackbody, g) single warm absorber fit with blackbody, h), same as g) but with Fe K$\alpha$ line.}
\label{Fits}
\end{figure*}

Both quasar spectra were fitted over the 0.3-10.0 keV energy range with a simple model, consisting of a power law and Galactic absorption. The parameters for these fits are shown in Table~\ref{Power law and warm absorber parameters}. Fig.~\ref{Fits} shows the spectra and data/model ratios for the combined {\sl pn} and {\sl MOS} data of each quasar. All spectra are plotted in the observed frame, unless otherwise stated. During the spectral fitting, neutral Galactic absorption was held fixed at the values shown in Table.~\ref{quasars}.

\section{Observations and Data Reduction}
\label{Observations}

PG~1114+445 and PG~1309+355 were observed with {\sl XMM-Newton} (Jansen et al 2001) as part of the {\sl OM} Guaranteed Time programme. Table~\ref{quasars} summarises the Galactic column densities (from Dickey \& Lockman 1990), the redshifts (from L97), and observation dates. The {\sl RGS} data do not contain enough counts for us to construct a useful spectrum for either quasar.

\begin{sidewaystable*}
Parameters found from the power law and warm absorber model fits for PG~1114+445. Errors are at 90 \% confidence ($\Delta \chi^{2} = 2.71$). $\xi$ is the ionisation parameter in units of erg cm s$^{-1}$. PL norm is the normalisation of the power law, in units of $10^{52}$ ph s$^{-1}$ keV$^{-1}$. EW is the equivalent width of the iron K$\alpha$ line. Prob$_{null}$ is the null hypothesis probability.\\
\begin{tabular}{lcccccccclc}
\hline
PG~1114+445  & N$_{\mbox{\sc h}}$  & log $\xi$  & $\Gamma$ & PL norm & Fe K$\alpha$ & Fe K$\alpha$ & Fe K$\alpha$ & $\chi^{2}$/d.o.f & Prob$_{null}$   \\
& ($10^{22}$cm$^{-2}$)&& & at 1 keV & Energy (keV) & FWHM (eV) & EW (eV)   \\
\hline
PL & --- &  --- & 1.16$\pm{0.01}$ & 2.79${\pm0.04}$  & --- & ---  & --- & 4475/832 & 0 \\

PL$\times$xabs & 1.63$^{+0.07}_{-0.05}$  & 1.50$\pm{0.03}$  & 1.77$\pm{0.02}$ & 8.28$\pm{0.24}$  & --- & --- &  ---& 924/830 & 0.01 \\

PL$\times$xabs$\times$xabs  & 4.21$^{+1.00}_{-0.70}$, 0.71${\pm0.08}$  & 2.49$\pm{0.09}$, 0.88$^{+0.11}_{-0.10}$ & 1.87$^{+0.04}_{-0.03}$ & 10.29$\pm{0.60}$ & --- & --- & --- &  834/828 & 0.44 \\  

(PL+Fe K$\alpha$)$\times$xabs$\times$xabs  & 5.30$^{+1.30}_{-1.00}$, 0.74$^{+0.08}_{-0.07}$  &  2.57$\pm{0.08}$, 0.83$^{+0.10}_{-0.07}$   &  1.92${\pm0.04}$  & 11.0$^{+0.80}_{-0.70}$  & 6.51$^{+0.04}_{-0.05}$ & 320$^{+110}_{-90}$  &  230$^{+90}_{-70}$   & 766/825 & 0.93   \\

\hline
\\
\end{tabular}
\begin{tabular}{lcccccccccclc}
PG~1309+355  & N$_{\mbox{\sc h}}$   & log $\xi$  & BB kT & BB Area & $\Gamma$ & PL norm & Fe K$\alpha$ & Fe K$\alpha$ & Fe K$\alpha$ & $\chi^{2}$/d.o.f & Prob$_{null}$   \\ 
&($10^{22}$cm$^{-2}$) && (keV) & (10$^{23}$ cm$^{2}$) && at 1 keV & Energy (keV) & FWHM (eV)  & EW (eV)  \\
\hline
PL & --- & --- & --- & --- & 1.96$\pm{0.02}$  & 4.75$\pm{0.07}$  & --- & --- & --- & 743/456 & 4.34$\times10^{-16}$ \\

PL$\times$xabs & 1.00$^{+0.20}_{-0.10}$  & 2.03$^{+0.09}_{-0.08}$ & --- & --- & 2.08$^{+0.02}_{-0.03}$  & 6.45$\pm{0.20}$  & --- & --- & --- & 378/454 & 1  \\

(PL+BB)$\times$xabs & 0.40$^{+0.14}_{-0.12}$  & 1.86$\pm{0.20}$ & 0.12$\pm{0.02}$  & 3.20$^{+3.80}_{-1.50}$ & 1.85$\pm{0.05}$ & 4.75$\pm{0.30}$  & --- & --- & --- & 324/453 & 1 \\

(PL+BB+Fe K$\alpha$)$\times$xabs & 0.42$^{+0.14}_{-0.13}$ & 1.87$^{+0.10}_{-0.20}$ & 0.12$\pm{0.02}$ & 3.04$\pm{0.60}$ & 1.87$^{+0.06}_{-0.05}$ & 4.84$^{+0.40}_{-0.30}$ & 6.42$^{+0.12}_{-0.09}$  & 230$^{+380}_{-230}$  & 130$^{+130}_{-80}$  &  314/450 & 1  \\

\hline
\end{tabular}
\label{Power law and warm absorber parameters}
\end{sidewaystable*}

A power law was a poor fit to the {\sl EPIC} data of PG~1114+445, with a highly unacceptable $\chi^2$/dof of 5.38 for 832 dof. Similarly, a power law fit to {\sl EPIC} data of PG~1309+355 yields an unacceptable $\chi^2$/dof of 1.63 for 456 dof. A warm absorber is indicated because at 0.7 keV there are only $\sim$ 0.6 times the number of counts predicted by the power law model for this quasar.

\subsection{Model fits incorporating warm absorbers}

Each quasar was fitted with a model including a warm absorber, as well as a power law and Galactic absorption. We characterise each warm absorber with an ionisation parameter, $\xi$ (Tarter, Tucker \& Salpeter 1969). The ionisation parameter is defined as $\xi = L/nr^{2}$ in erg cm $s^{-1}$. Here $L$ is the ionising X-ray luminosity (erg s$^{-1}$), $n$ the gas density (cm$^{-3}$), and $r$ the distance of the ionising source from the absorbing gas, in cm. 

For the warm absorbers, we used the {\sl xabs} model in {\sl SPEX}, which applies line and photoelectric absorption by a column of photoionised gas. The ionisation parameters and column densities in {\sl xabs} were allowed to vary freely, but we assumed solar elemental abundances for the ions in the absorbers. 

We assumed that the velocity structures of the X-ray and UV absorbers in these quasars are the same. However, this is not necessarily true, for we show later that the absorbers could lie at different distances from the continuum source. In the models we used the UV absorption line outflow velocities and velocity dispersions from M98 for PG~1114+445, and from Bechtold, Dobrzycki \& Wilden (2002; hereafter BO2) for PG~1309+355. The velocity dispersions were found using the relation $\sigma$ = FWHM/2.35.

\begin{figure*}
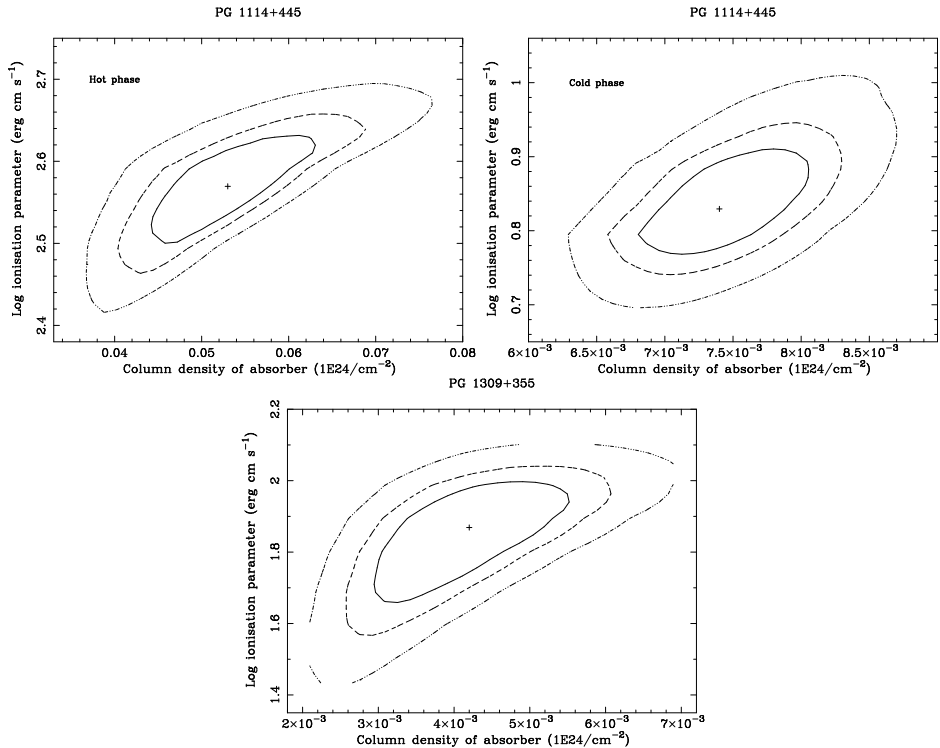

\epsfig{file=fig2.eps,width=4.9cm,angle=270}
\epsfig{file=fig3.eps,width=4.9cm,angle=270}
\epsfig{file=fig4.eps,width=4.9cm,angle=270}
\caption{Top: Confidence contours for each warm absorber in PG~1114+445. Left, higher-ionisation phase; right, lower-ionisation phase. Bottom: Confidence contours for the single warm absorber in PG~1309+355. Each contour denotes 68\%, 90\% and 99\% confidence, moving from the centre out.}
\label{Contour plots}
\end{figure*}

\subsection{PG~1114+445}

PG~1114+445 was first modelled with one warm absorber, which greatly improves the fit, with $\chi^2$ decreasing by 3551 for 2 extra parameters. However, from examination of the data/model ratio in Fig.~\ref{Fits}, there are more absorption components present than are accounted for by a single warm absorber of log $\xi$ of 1.5; in the observed frame, absorption features are noticeable at $\sim$ 0.7 keV, possibly due to an iron UTA, and $\sim$ 0.8-0.9 keV, probably due to L-shell iron. The single warm absorber has too high an ionisation parameter to account for these features, and implies another ionisation phase is present.

To account for these features, we tried adding a second warm absorber phase, which improved the $\chi^2$ by 90 for 2 additional free parameters. The two absorbers in the model have very different values of log $\xi$; the high-ionisation (`hot') phase has log $\xi$ of 2.49, N$_{\mbox{\sc h}}$ = 4.2$\times10^{22}$ cm$^{-2}$, and the low-ionisation (`cold') phase of warm absorber has log $\xi$ of 0.88, N$_{\mbox{\sc h}}$ = 0.71$\times10^{22}$ cm$^{-2}$. The cooler phase accounts for the Fe UTA, and the hotter phase accounts for the L-shell iron, see Section.~\ref{Discussion}. The details of the warm absorber models are given in Table~\ref{Power law and warm absorber parameters}. A 6.5 keV (rest-frame) iron K$\alpha$ line, is prominent in the spectrum; this is incorporated into the model, for plot d) of Fig.~\ref{Fits}. We find the equivalent width to be 230$^{+90}_{-70}$ eV, whereas George et al (1997) find an equivalent width of 60$^{+120}_{-60}$ eV. Having added this line to the model, the parameters of the warm absorbers do not change significantly; the hot phase has log $\xi$ of 2.57, N$_{\mbox{\sc h}}$ = 5.3$\times10^{22}$ cm$^{-2}$, and the low-ionisation (`cold') phase of warm absorber has log $\xi$ of 0.83, N$_{\mbox{\sc h}}$ = 0.74$\times10^{22}$ cm$^{-2}$. The confidence intervals on the ranges of $\xi$ and column density for each phase are shown in Fig.~\ref{Contour plots}. To investigate whether more warm absorber phases are present, we tried adding a third warm absorber to the model. However this extra phase only improved the fit by a $\Delta \chi^2$ of 17, and so we have not investigated it further.  

In Fig.~\ref{Fits} b)-d) the very soft part of the spectrum does not appear to be fitted particularly well; this may indicate some small amount of neutral absorption is present. However the reduced chi-squared is already less than 1, therefore an additional model component is not warranted.

The 0.3-10 keV flux of the final model is 3.1$\times10^{-12}$ erg s$^{-1}$ cm$^{-2}$, compared to 1.7$\times10^{-12}$ erg s$^{-1}$ cm$^{-2}$ (estimated using PIMMS) from the {\sl ASCA} observation of George et al (1997).

\subsection{PG~1309+355}

A power law and a single warm absorber are a reasonable fit for PG~1309+355, as shown in Fig.~\ref{Fits}. We added a soft excess blackbody component to the model, which reduced the $\chi^2$ by 54 for 2 extra free parameters. The best-fitting model has log $\xi$ = 1.87, and a blackbody temperature of 0.12 keV. An iron K$\alpha$ line at 6.4 keV (rest-frame), with equivalent width 130$^{+130}_{-80}$ eV, is detected; plot h) of Fig.~\ref{Fits} includes this line in the model. The fit parameters are given in Table~\ref{Power law and warm absorber parameters}. From the best-fitting model, we produced confidence intervals for $\xi$ and column density, and these are shown in Fig.~\ref{Contour plots}. The 0.3-10 keV flux of the final model is 1.3$\times10^{-12}$ erg s$^{-1}$ cm$^{-2}$.

\section{Discussion}
\label{Discussion}

\begin{figure*}
\epsfig{file=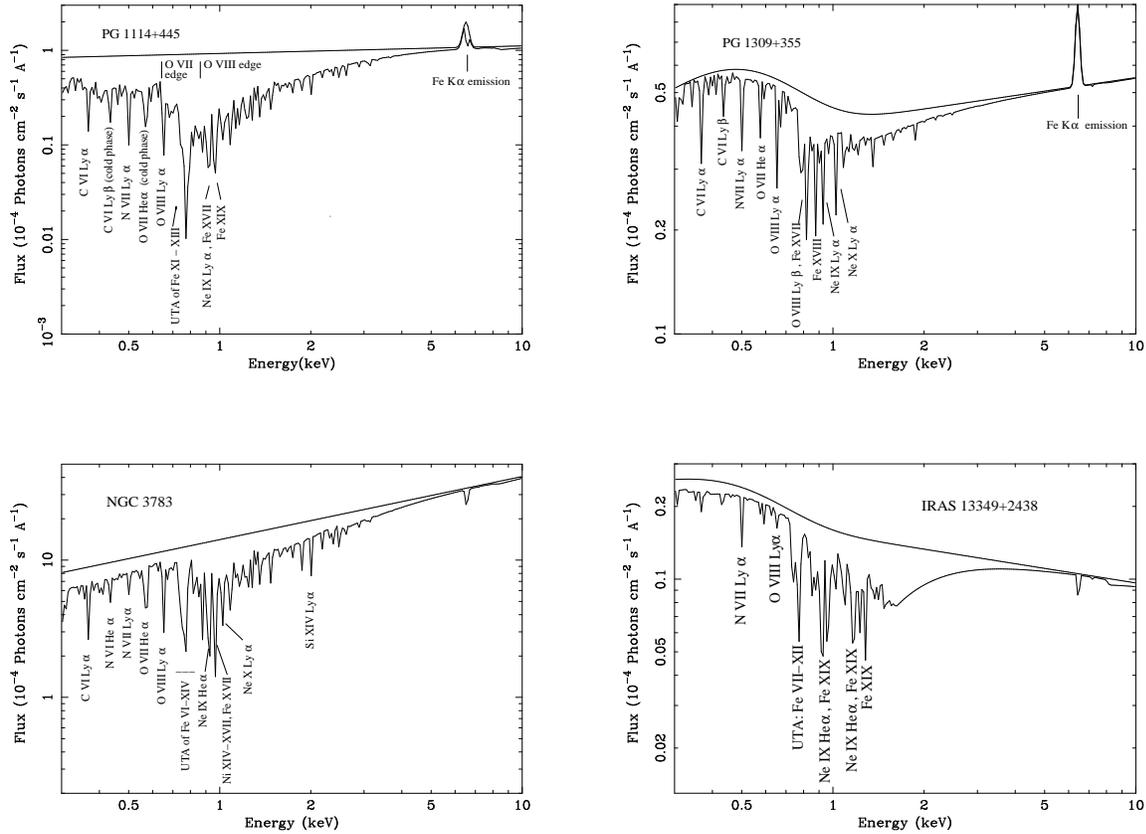,height=11cm,angle=0}
\caption{Models of the warm absorbers plotted at high resolution, in the rest frame. The dominant ions are labelled. Clockwise from top left: PG~1114+445, PG~1309+355, IRAS 13349+2438 (Sako et al 2001), NGC 3783 (Blustin et al 2002). For comparison, the solid lines show the models with no ionised absorption.}
\label{RGS models}
\end{figure*}

We have obtained useful constraints on the warm absorbers in the quasars PG~1114+445 and PG~1309+355, from a detailed analysis. In order to determine the main absorbing species and see how they compare with warm absorbers observed in other AGN, the best-fitting models are plotted at a resolution higher than that of {\sl EPIC}, in Fig.~\ref{RGS models}. The main ionic species contributing to the warm absorption in each AGN are labelled. These models do not include Galactic absorption, and are shown in the rest frames of the AGN. 

For PG~1114+445, the most important ions in the cold phase are O V-VII, and Fe XI-XIII. The iron UTA, an indication of a low-$\xi$ phase in a warm absorber, is very prominent in the model of PG~1114+445; the UTA consists of iron ions towards the top of the M-shell range, encompassing Fe IX-XIV. The high-$\xi$ phase is dominated by the ions O VIII and Fe XVIII-XXII. The absorber in PG~1309+355 is too highly ionised for there to be an iron UTA; the most important ions in the single phase absorber of this quasar are O VII-VIII and Fe XIII-XVIII.

P04 model {\sl EPIC} data of PG~1114+445 and PG~1309+355 with simple absorption edges of O VII and O VIII. For PG~1114+445, they obtain optical depths of 2.26$^{+0.22}_{-0.19}$ for the OVII edge, and 0.32$\pm{0.16}$ for the OVIII edge. For PG~1309+355 they obtain 0.46$^{+0.20}_{-0.15}$ for the OVII edge, and $<$ 0.25 for the OVIII edge. 

Converting from the optical depth values of P04, for the warm absorber of PG~1114+445, assuming a single phase of absorption we find that the optical depths correspond to an equivalent hydrogen column of 2$\times10^{22}$ cm$^{-2}$. Similarly, for PG~1309+355, the optical depth corresponds to an equivalent hydrogen column density of 7$\times10^{21}$ cm$^{-2}$.

From our detailed fitting, we obtain columns of 5.3$\times10^{22}$ cm$^{-2}$ and 7.4$\times10^{21}$ cm$^{-2}$ for the two absorbers in PG~1114+445, and 4.2$\times10^{21}$ cm$^{-2}$ for the single absorber PG~1309+355; these are similar to the converted values of Porquet et al (2004), again taking into account the estimation uncertainties.

\subsection{Distances, filling factors and densities of the absorbers}
\label{Distance and densities of the absorbers}

The distance, $r$, to the warm absorber from the continuum source has been speculated to range from 10$^{17}$ to 10$^{19}$ cm (Krolik \& Kriss 2001). Assuming $\delta r/r \leq 1$, i.e. the depth of the absorber cannot be greater than its distance from the source, using the relation $\xi = L/nr^{2}$ for the ionisation parameter, and taking N$_{\mbox{\sc h}} = n \delta r f$, where $f$ is the filling factor of the gas, we obtain  
\begin{equation}
r \leq \frac{L_{ion}f}{{\rm N_{H}} \xi} 
\label{equation1}
\end{equation}
We assume the conservative upper limit of $f \leq 1$. We can therefore calculate upper limits for the distances from the continuum source to the warm absorber. This yields a distance of $\leq 3.9\times10^{19}$ cm to the higher-$\xi$ absorber in PG~1114+445, and $\leq 1.4\times10^{22}$ cm to the lower-$\xi$ absorber. For PG~1309+355, we calculate an upper limit on the absorber distance of 1.9$\times10^{21}$ cm.

Assuming that the outflow velocity, derived from the UV data, exceeds the escape velocity, which is defined as 
\begin{equation}
v_{escape} = \sqrt{\frac{2GM}{r}} 
\label{equation4}
\end{equation}
where $M$ is the black hole mass, we can find a lower limit $r_{min}$ on the distance from the absorber to the continuum source:
\begin{equation}
r \geq r_{min} = \frac{2GM}{v_{outflow}^2} 
\label{equation7}
\end{equation}
Using the black hole masses given in Table~\ref{correlations}, we find that $r \geq 2.4\times10^{19}$ cm for PG~1114+445. Similarly, for PG~1309+355, we find that the minimum distance of the absorber from the continuum source is 1.9$\times10^{18}$ cm.

Substituting $r_{min}$ for $r$ in equation~\ref{equation1}, we can find a lower limit for the filling factor:
\begin{equation}
f \geq \frac{r_{min} {\rm N_{H} \xi}}{L_{ion}}
\label{equation8}
\end{equation}
In this way we obtain $f \geq $ 0.62 for the higher-$\xi$ absorber of PG~1114+445, and $f \geq $ 1.8$\times10^{-3}$ for the lower-$\xi$ absorber. For PG~1309+355 we find that $f \geq $ 1.0$\times10^{-3}$.

To compute lower limits on the gas density, we use the relation $n$ = N$_{\mbox{\sc h}}/\delta r f$ with the maximum and minimum values of $r$, N$_{\mbox{\sc h}}$ and $f$; for lower bounds on $n$ we assume $f \leq 1$. For PG~1114+445, we find 1100 cm$^{-3}  \leq n \leq $ 4400 cm$^{-3}$ for the higher-$\xi$ phase, and 0.5 cm$^{-3}  \leq n \leq 2.0\times10^{5}$ cm$^{-3}$ for the lower-$\xi$ phase. We obtain limits of 1.6 $\leq n \leq $2.9$\times10^{6}$ cm$^{-3}$ for PG~1309+355.

In PG~1114+445, it is possible that there is a range of log $\xi$ in the absorber, instead of two discrete clouds. The log $\xi$ values in the best-fitting model would then reflect the dominant ionisation states within the warm absorber. A range of log $\xi$ has been observed in several objects, e.g. NGC 3783 (Blustin et al 2002); NGC 5548 (Steenbrugge et al 2003); IRAS 13349+2438 (Sako et al 2001).

\subsection{Outflow and accretion rates}
\label{Outflow and accretion rates}

In order to investigate the dynamics of the quasars we calculated the mass outflow rate, ${\dot{M}}_{out}$. Assuming a spherical outflow, we obtain:
\begin{equation}
{\dot{M}}_{out}  = 4\pi v m_{p} n r^2   
\label{equation11}
\end{equation}
where $v$ is the outflow velocity, $m_{p}$ is the proton mass, $n$ is the density and $r$ is the absorber distance from the continuum source. We assume the velocity derived from the UV data for the outflow velocity. Substituting $\xi = L/nr^{2}$ into equation~\ref{equation11}, we derive:
\begin{equation}
{\dot{M}}_{out}  = \frac{\Omega v m_{p} L_{ion}f}{\xi}  
\label{equation2}
\end{equation}
where $\Omega$ is the solid angle subtended by the outflow, and $f$ is the filling factor. We assume the warm absorber fills all lines of sight not blocked by the molecular torus, and therefore take the solid angle of the warm absorber to be 0.2$\times4\pi$, based on the observed ratio of Seyfert 1s to Seyfert 2s (Maiolino \& Rieke 1995). 

For the hot phase in PG~1114+445, we calculate that 2.6 M$_{\odot}$ yr$^{-1} \leq {\dot{M}}_{out} \leq 6.0 $M$_{\odot} $yr$^{-1}$, and in the cold phase, 0.4 M$_{\odot}$ yr$^{-1} \leq {\dot{M}}_{out}  \leq 320 $M$_{\odot}$ yr$^{-1}$. For PG~1309+355 we calculate a mass outflow of 0.03 M$_{\odot}$ yr$^{-1} \leq {\dot{M}}_{out} \leq 54 $M$_{\odot}$ yr$^{-1}$. These mass outflows for the absorbers can be compared with the mass accretion rates for each quasar,
\begin{equation}
{\dot{M}}_{acc} = \frac{L_{bol}}{\eta c^{2}}
\label{equation3}
\end{equation}
where $L_{bol}$ is the bolometric luminosity, and $\eta$ is the accretion efficiency, which we assume to be 0.1. For PG~1114+445, $L_{bol} = 5\times10^{45}$ erg s$^{-1}$ (M98), and for PG~1309+355, $L_{bol} = 6.8\times10^{45}$ erg s$^{-1}$ (Laor 1998). For PG~1114+445, we obtain a mass accretion rate of 0.9 M$_{\odot}$ yr$^{-1}$. Therefore, the outflow is similar to, or larger than, the accretion rate in PG 1114+445. The accretion rate of PG~1309+355 is 1.2 M$_{\odot}$ yr$^{-1}$.

The kinetic luminosity carried in the outflows for the absorbers was found using the relation 
\begin{equation}
L_{KE} = \frac{\dot{M}_{out} v^2 f}{2}
\label{equation5}
\end{equation}
For the higher-$\xi$ absorber in PG~1114+445, we find 2.3$\times10^{41}$ erg s$^{-1} \leq L_{KE} \leq 5.3\times10^{41}$ erg s$^{-1}$, while for the lower-$\xi$ absorber, we find 3.4$\times10^{40}$ erg s$^{-1} \leq L_{KE} \leq 2.9\times10^{43}$ erg s$^{-1}$; the two warm absorber phases together account for $10^{-3}$-$10^{-5}$ of the bolometric luminosity. For PG~1309+355, we find 1.8$\times10^{40}$ erg s$^{-1} \leq L_{KE} \leq 3.8\times10^{43}$ erg s$^{-1}$; so for this quasar, the kinetic luminosity of the outflow is $10^{-3}$-$10^{-6}$ times the bolometric luminosity.

\subsection{Associated UV and X-ray absorber?}
\label{Associated UV and X-ray absorber?}

As strong evidence exists for an association between X-ray and UV absorption systems, (e.g. M98, Crenshaw et al 1999), we investigated which phases of the X-ray absorber are related to the UV absorbers in the two PG quasars. We used the {\sl SPEX xabs} model to compute the equivalent widths of the UV absorption lines, due to the warm absorbers modelled in the X-ray data. These are shown in Table~\ref{Associate}.

\begin{table*}
\caption{X-ray and UV absorber line equivalent widths in \AA. B02 refers to Bechtold (2002).}
\begin{tabular}{ccccccccc}
\hline
Line, \AA& PG~1114+445 & PG~1114+445  &  PG~1114+445  & PG~1309+355  &  PG~1309+355   \\
&  model: log$\xi=0.88$  &  model: log$\xi=2.49$  & observed (M98) &  model: log$\xi=1.86$  & observed (BO2) \\
\hline

C IV, 1548.2         &  5.5      &     ---         &  2.4	&  0.02	     & 1.63	\\
C IV, 1550.8         &  4.8      &     ---         &  2.1	&  0.01      & 0.43	\\

N V, 1238.8          &  4.8      &     ---	   &  1.8	&  0.09      & 1.85	\\
N V, 1242.8          &  4.4      &     ---         &  1.9	&  0.04      & 1.66	\\

O VI, 1037.6         &  5.2      &	0.06	&  ---	&  1.8       & 1.99	\\

\hline
\end{tabular}\\
\label{Associate}
\end{table*}

The UV lines observed by M98 all appear in the low-$\xi$ phase model of PG~1114+445, with comparable equivalent widths, so we identify the UV absorber with the low-$\xi$ X-ray absorber phase. For the low-$\xi$ absorber, our assumption that the velocity structure is the same as that of the UV absorber, is therefore verified. However, the high-$\xi$ X-ray absorber is too highly ionised to give rise to any of the UV lines observed by M98.

For the case of PG~1309+355, the X-ray absorber is not predicted to give rise to significant C IV or N V lines, as it is too highly ionised. However, the X-ray absorber model does predict absorption from O VI at 1037 \AA\ with an equivalent width similar to that observed. Therefore it is likely that the X-ray and UV absorbers have the same velocity structure.

\subsection{Where is the warm absorber coming from?}
\label{Where is the warm absorber coming from?}

In order to locate the warm absorbers in relation to the structures of the AGN considered here, we first have to calculate the locations of the Broad Line Region (BLR) and the torus in each quasar. 

The quasars considered here are Type 1 AGN so that we are looking directly towards the central engine. By finding the locations of the warm absorbers relative to the BLR and the molecular torus, we have a useful picture of the structure of each quasar, and we can constrain where the warm absorbers come from.

We can estimate the BLR sizes using the empirical relation of Equation 6 in Kaspi et al (2000), derived from reverberation studies:
\begin{equation}
R_{\rm BLR}  = (32.9^{+2.0}_{-1.9})\left[\frac{\lambda L_{\lambda}(5100 \mbox{\rm \AA})}{10^{44} {\rm erg s^{-1}}}\right]^{0.700\pm0.033}  {\rm lt-days}
\label{equation6}
\end{equation}
In this equation, $\lambda L_{\lambda} (5100 \mbox{\rm \AA})$ is the monochromatic luminosity of the continuum at 5100 \AA. We use the $\lambda L_{\lambda} (5100 \mbox{\rm \AA})$ values of 3.9$\times 10^{44}$ erg s$^{-1}$ for PG~1114+445, and 6.9$\times 10^{44}$ erg s$^{-1}$ for PG~1309+355, from Vestergaard (2002). We calculate the BLR distance in PG~1114+445 to be 2.1$\times10^{17}$ cm. We use the relation in Krolik \& Kriss (2001; see also Barvainis 1987) to estimate the inner edge of the torus from the ionising luminosity $L_{ion,44}$, in units of 10$^{44}$ erg s$^{-1}$:
\begin{equation}
R_{\rm torus}  \sim L^{1/2}_{ion,44}  
\label{equation9}
\end{equation}
We estimate the torus inner edge to be 2.3$\times10^{18}$ cm for PG~1114+445. Therefore, the absorbers in this AGN are located at a much greater distance from the continuum source than the BLR, and somewhat further than the inner edge of the torus.

A similar situation exists for PG~1309+355, where the distance to the BLR is 3.3$\times10^{17}$ cm, and the torus inner edge is at 1.6$\times10^{18}$ cm. As the absorber lies further than 1.9$\times10^{18}$ cm, the warm absorber is again much further than the BLR and at least as far as the torus inner edge.    

Since the warm absorbers appear to lie beyond the inner edge of the torus, a torus wind is their most likely origin in these quasars. At these distances, the warm absorbers are within the narrow line region (NLR). Similarly, Sako et al (2000) suggested this was the case for IRAS 13349+2438, while in NGC 1068 the warm emitter is directly observed to be in the NLR (Ogle et al 2002, Brinkmann et al 2003.)

An alternative possibility for the origin of warm absorbers is a wind launched from the accretion disc. This is modelled by Proga, Stone \& Kallman (2000), who consider a disc accreting onto a 10$^{8}$ M$_{\odot}$ black hole. They find that the disc radiation can launch a wind at $\sim$ 10$^{16}$ cm from the central engine, and can accelerate it up to 15,000 km s$^{-1}$ at an angle of $\sim$ 75 degrees to the disc axis. This can be considered for both the quasars in this paper, which have black hole masses of $\sim$ 10$^{8}$ M$_{\odot}$ (Vestergaard 2002). The model has a mass loss of 0.5 M$_{\odot}$ yr$^{-1}$ and typical column densities generated in the wind are $\sim$ 10$^{23}$ cm$^{-2}$, at least an order of magnitude higher than the warm absorbers we observe in PG~1114+445 and PG~1309+355. Furthermore, the model predicts that the density of the outflow is greatest near the base of the flow, at $\sim$ 10$^{16}$ cm, and therefore if an accretion disc wind were responsible for the warm absorbers, we would expect to observe them at this distance from the continuum source. However, we observe them at $ > 10^{18}$ cm in both quasars. Thus in the quasars considered here, the warm absorbers lie at distances too great to be produced by an accretion disc wind as described by Proga et al (2000). 

There have been recent reports of high-energy outflows in the quasars PG~1211+143 (Pounds et al 2003) and PDS~456 (Reeves, O'Brien \& Ward 2003), however these quasars were both found to exhibit very high ionisation parameters and high columns. PG~1211+143 was found to have log $\xi$ of 3.4, a high column of 5 x 10$^{23}$ cm$^{-2}$ and an outflow of 0.12 M$_{\odot}$ yr$^{-1}$ at a velocity of 0.1c. Similarly, for PDS~456, a log $\xi$ of 2.5 and a column of 5 x 10$^{23}$ cm$^{-2}$ are found for a mass outflow, with a rate of 10 M$_{\odot}$ yr$^{-1}$ at a velocity of 0.2c. It is interesting that the log $\xi$ of 2.5 for PDS~456 is so similar to the log $\xi$ of 2.57 found for one of the absorbers in PG~1114+445. However, these outflows in PG~1211+143 and PDS~456 are much faster than those we consider in this paper. Also, we have assumed the velocities of the X-ray absorbers are the same as those of the UV absorbers for PG~1114+445 and PG~1309+355.

\subsection{Comparison with other AGNs}
\label{Comparison with other AGNs}

It is intriguing that the best-fitting model parameters of PG~1114+445 ($L_{bol}$ = 5$\times10^{45}$ erg s$^{-1}$), shown in Table~\ref{Power law and warm absorber parameters}, are similar to those of NGC 3783 ($L_{bol}$ = 4.5$\times10^{44}$ erg s$^{-1}$; Markowitz et al 2003), the nearest Seyfert 1 AGN with a warm absorber which has been well studied (see e.g. Kaspi et al 2002; Blustin et al 2002; Netzer et al 2003; Krongold et al 2003). In PG~1114+445, we find values of log $\xi$ = 2.57 and log $\xi$ = 0.83 for the higher- and lower-ionisation absorbers respectively. Blustin et al (2002) model NGC 3783 with two warm absorbers, a higher-ionisation one with log $\xi$ = 2.4, N$_{\mbox{\sc h}}$ $\sim 10^{22}$ cm$^{-2}$, and a lower-ionisation one with log $\xi$ = 0.3 and N$_{\mbox{\sc h}} \sim 10^{20}$ cm$^{-2}$. PG~1114+445 is ten times higher in luminosity than NGC 3783, yet the ionisation parameters and column densities of their absorbers are comparable. It appears that the ionisation parameters of warm absorbers in at least some quasars are similar to those in Seyfert galaxies. George et al (1997) also noted that the warm absorber in PG~1114+445 is akin to those found in Seyfert galaxies.

Table~\ref{correlations} shows a comparison of warm absorber parameters in the two quasars studied here, along with NGC 3783 and IRAS 13349+2438. The ionisation parameters and column densities of the absorbers do not show any particular trend with luminosity. This supports the hypothesis that there is little difference in AGN warm absorber characteristics across a sizeable range in $L_{bol}$. From the definition of the ionisation parameter $\xi = L/nr^{2}$, and given that similar values of $\xi$ are found over a large range of luminosity, we can infer that $r$ scales approximately with $L^{1/2}_{ion,44}$. Since $r$ has the same luminosity dependance as the torus distance in Equation~\ref{equation9}, this result is consistent with the warm absorber originating as a torus wind. However, the distances inferred for the warm absorbers are not sufficently constrained in Table~\ref{correlations} to be able to construct a global picture of where a warm absorber would lie in an AGN structure, given its bolometric luminosity.

\begin{table*}
\caption{AGN warm absorber parameters. The objects are listed in increasing order of bolometric luminosity, L$_{bol}$. Log $\xi$ is the ionisation parameter.}
\begin{tabular}{ccccccccccccc}
\hline
AGN && M$_{BH}$/M$_{\odot}$$^{a}$ & $L_{bol}$$^{b}$ & log $\xi$  & log N$_{\mbox{\sc h}}$  & Outflow vel & W.A. distance \\
&&& erg s$^{-1}$ & erg cm s$^{-1}$  & cm$^{-2}$  & km s$^{-1}$ & 10$^{18}$ cm \\
\hline
NGC 3783$^{c}$     & Sy 1  & 9.3$\times10^{6}$    &  2.6$\times10^{44}$       & 0.3, 2.4          & 20.73, 22.45  & 800, 800 &  0.17-42$\times10^{3}$, 0.17-6.5$^{g}$  \\


PG~1114+445  & QSO  & 2.6$\times10^{8}$       &  5.0$\times10^{45}$       & 0.83, 2.57        & 21.87, 22.72   & 530, 530(?)$^{d}$ & 24.3-8$\times10^{3}$, 24.3-33.0 \\

PG~1309+355  & QSO  & 1.6$\times10^{8}$       &  6.8$\times10^{45}$       & 1.87              & 21.62   & 213 & 1.9-750   \\

IRAS 13349+2438$^{e}$  & QSO  &  8$\times10^{8}$     &  8.4$\times10^{45}$       & 0, 2-2.5    & 21.30, 22.40$^{f}$  & 420, 0  & 39.0-2$\times10^{5}$, ?$^{h}$ - 102.0$^{g}$  \\ 
\hline
\end{tabular}\\
\begin{flushleft}
$^{a}$ M$_{BH}$/M$_{\odot}$ values for NGC 3783, PG~1114+445 and PG~1309+355 from Vestergaard (2002); IRAS 13349+2438 from Brandt et al (1997). \\

$^{b}$ L$_{bol}$ values for NGC 3783 from Woo \& Urry (2002); PG~1114+445 from M98; PG~1309+355 from Laor (1998); IRAS 13349+2438 from Beichman et al (1980)   \\

$^{c}$ Blustin et al (2002) \\

$^{d}$ The velocity of the higher-ionisation absorber is less certain, as it is not expected to contribute to the UV absorption lines. \\

$^{e}$ Sako et al (2001) \\

$^{f}$ Average columns  \\

$^{g}$ Blustin et al (2004), in preparation.  \\

$^{h}$ Blustin et al (2004), in preparation: the phase with log $\xi$ = 2.25 has an outflow speed of zero, so a minimum distance cannot be found for this absorber. \\

\end{flushleft}

\label{correlations}
\end{table*}

\subsection{Conclusions}
\label{Conclusions}

We have studied the warm absorption in the quasars PG~1114+445 and PG~1309+355 using {\sl XMM-Newton EPIC} observations. The absorption in PG~1114+445 consists of at least two phases, a `hot' phase with a log ionisation parameter $\xi$ of 2.57, and a `cooler' phase with log $\xi$ of 0.83. The most important ions in the cold phase are O V-VII and Fe XI-XIII, whilst for the high-$\xi$ phase, the main ions are O VIII and Fe XVIII-XXII. We observed a UTA of M-shell iron in the cooler phase of the absorber. We note that this quasar exhibits absorption which is similar to that observed in the Seyfert 1 NGC 3783. We can model the absorption in PG~1309+355 with a single warm absorber phase, which has log $\xi$ of 1.87; the most important ions in this absorber are O VII-VIII and Fe XIII-XVIII. We find that the absorbers are located at distances of 10$^{18}$ - 10$^{22}$ cm from the continuum sources in these AGN; these distances suggest that the absorbers are winds from molecular tori, rather than accretion disc winds. The distances to the warm absorbers from the central continuum source are proportional to the square root of the AGN ionising luminosity, which is consistent with a torus wind origin. The kinetic luminosities of the outflows are insignificant compared to the bolometric luminosities of the quasars.

\section*{Acknowledgments}
\label{Acknowledgments}

CEA thanks Rhaana Starling, Roberto Soria and Catherine Brocksopp for useful discussions, and financial support from a PPARC quota studentship. This work is based on observations obtained with XMM-Newton, an ESA science mission with instruments and contributions directly funded by ESA Member States and the USA (NASA).

\end{document}